\documentclass[RNAAS]{aastex62}
\usepackage{url}
\usepackage{natbib}
\newcommand\lb{LiteBIRD }

\begin{document}

\title{Simulating Calibration and Beam  Systematics for future CMB space  mission with TOAST package}

\correspondingauthor{Giuseppe Puglisi}
\email{gpuglisi@berkeley.edu}

\author{Giuseppe Puglisi}
\affiliation{Computational Cosmology Center, Lawrence Berkeley National Laboratory, Berkeley, CA 94720, USA }
\affiliation{Space Sciences Laboratory at University of California, 7 Gauss Way, Berkeley, CA 94720 }
\affiliation{Department of Physics, University of California, Berkeley, CA, USA 94720 }
\author{Reijo Keskitalo }
\affiliation{Computational Cosmology Center, Lawrence Berkeley National Laboratory, Berkeley, CA 94720, USA }
\affiliation{Space Sciences Laboratory at University of California, 7 Gauss Way, Berkeley, CA 94720 }
\affiliation{Department of Physics, University of California, Berkeley, CA, USA 94720 }
\author{Ted Kisner }
\affiliation{Computational Cosmology Center, Lawrence Berkeley National Laboratory, Berkeley, CA 94720, USA }
\affiliation{Space Sciences Laboratory at University of California, 7 Gauss Way, Berkeley, CA 94720 }
\affiliation{Department of Physics, University of California, Berkeley, CA, USA 94720 }
\author{Julian D. Borrill }
\affiliation{Computational Cosmology Center, Lawrence Berkeley National Laboratory, Berkeley, CA 94720, USA }
\affiliation{Space Sciences Laboratory at University of California, 7 Gauss Way, Berkeley, CA 94720 }
\affiliation{Department of Physics, University of California, Berkeley, CA, USA 94720 }


\keywords{cosmology, cosmic microwave background, systematics uncertainties, simulations, surveys}

\begin{abstract}
    We address in this work the   instrumental systematic errors  that can potentially  affect the forthcoming and future Cosmic Microwave Background experiments aimed at observing its polarized emission. In particular, we focus on the systematics induced by the  beam and calibration, which are considered the major sources of leakage from  total intensity measurements to  polarization. We simulated synthetic data sets with TOAST,  a  publicly available simulation and data analysis  package.  We also propose a mitigation technique aiming at reducing the  leakage by means of a template fitting approach.  This technique has shown promising results reducing the leakage by 2 orders of magnitude at the power spectrum level when applied to a realistic simulated data set of the \lb  satellite mission. 
     
\end{abstract}
\section{Introduction}
The Cosmic Microwave Background (CMB) emission provides one of the most favorite channels for probing the Universe at large scales.   In particular, CMB forthcoming experiments are increasingly focusing on measuring    the primordial CMB polarization  B-mode ( expected to peak at the degree scales \citep{1997PhRvL..78.2054S,Hu:1997hv}) which is directly linked to a stochastic gravitational wave background emitted at the time of {inflation}\citep{1981PhRvD..23..347G,1982PhLB..117..175S}.    The amplitude of the primordial B-modes is quantified by the {tensor-to-scalar ratio}, $r$. To date,   the best constraints on $r$  have recently been set to $r<0.07$  by \citet{bk2018,bicepkeck2020demonstration}.

A critical piece of this framework will necessarily be the ability of generating synthetic mission data sets of sufficient realism, both in their complexity and their size, to be truly representative of the data set that would be gathered by a given mission configuration. In this research note, we aim at  showing preliminary results of simulations encoding systematic effects injected with the Time-Ordered Astrophysics Scalable Tools (TOAST) package, namely  calibration errors,  gain fluctuations and optical beam  asymmetries.

\begin{figure} 
\begin{center}
\includegraphics[scale=0.4,angle=0]{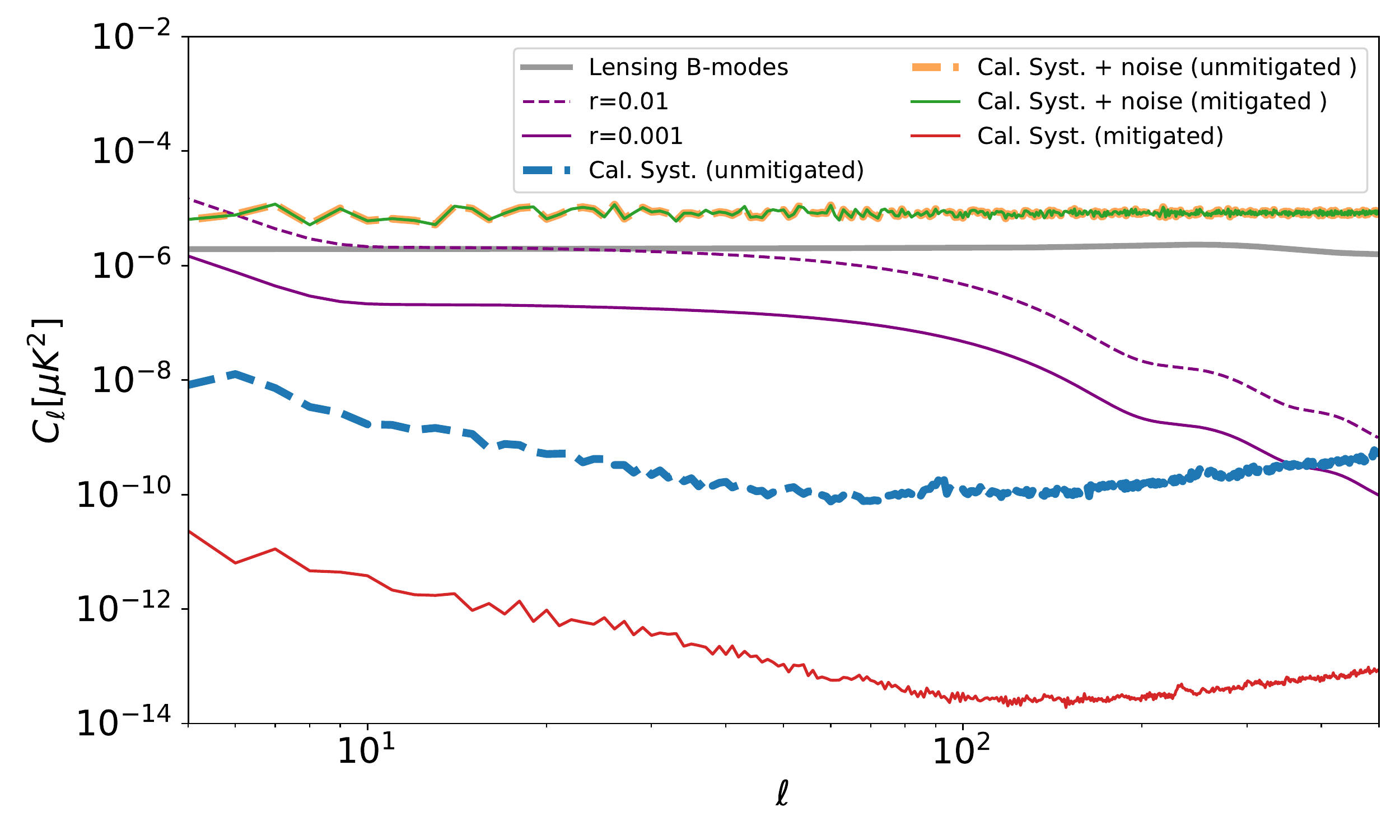}
\caption{ Angula power spectra of B-modes. Spectra obtained from the residual maps encoding  noiseless simulations with (dashed thick blue) calibration systematics and  (solid red)  mitigated systematics with the template fitting procedure outlined in Section\ref{sec:mitig}. Spectra obtained with noisy simulations encoding   no-mitigation    and mitigation cases are shown  respectively in (dashed thick orange ) and (solid green). As a reference, we include the theoretical angular spectra for  primordial  $r=0.01$ (dashed  thin purple  ) and $r=0.001$ (solid thin purple) and from the gravitational lensing B-modes (thick solid grey). }\label{fig:1}
\end{center}
\end{figure}

\section{Calibration systematics } \label{sec:calerr }
We consider two source of calibration systematics: calibration uncertainties and a long term fluctuation of the calibration gain (commonly referred as  \emph{gain drift}) due to thermal instabilities in the focal-plane. The former essentially quantifies the  error affecting the calibration measurements performed between two consecutive observations, e.g. for the \lb  satellite, \citep{Sugai_2020}, we assume the calibration to happen every 24 hours.  Calibration errors can be thus simulated by drawing Gaussian random values centered around $1$ and with a   width (hereafter assumed to be $1 \%$). 

We also simulate the instrumental thermal instability which can lead  into changes of the gain calibration during one observation time scale, injected as  a slowly varying drift  with a $1/f$ frequency spectrum, typical for the CMB measurements.  Moreover, the amplitude of the drift fluctuations   can be directly linked to an effective temperature related to  fluctuations in the focal-plane unit. 

The gain drift signal is parametrized by  the following power spectrum density: 
\begin{displaymath} 
PSD(f) = \delta_g \left( \frac{f_{knee}}{f} \right) ^{\alpha},
\end{displaymath}
with $\alpha=1$, $f_{knee}=20 $ mHz, $\delta_g= 10 \mu \mathrm{K} $.

Both  calibration errors and  gain drift are simulated independently for each  detector and   observation.
 
 \subsection{Template fitting mitigation}\label{sec:mitig}
 
 In order to mitigate the systematic residuals induced by   calibration systematics described in the previous Section, we outline below a technique aimed at further reduce the leakage. 
 
 We assume that both  calibration errors and gain drifts   can be approximated  as a linear combination of \emph{Legendre polynomials} up to a certain order, $n_{poly} \lesssim 4 $, within each observation: 
 
 \begin{equation}
     g(t) = \sum _i ^{n_{poly}} \hat{a}_i \mathcal{L } ^{(i) }(t) .\label{eq:expansion }
 \end{equation}
 
 The template fitting mitigation proposed in this study consists in estimating the weights $\hat{a} $  that minimize a $\chi^2$ problem provided we have an approximated estimate of the underlying signal via a \emph{template map} (a similar approach can be found in \citep{npipe2020,Keihanen2010}): 
 \begin{equation}
 (F^t C_n^{-1} ZF) \hat{a}  = F^t C_n^{-1} Z d, \label{eq:pcg}
 \end{equation}
  with $C_n$ being the noise covariance matrix, $F$  the matrix built from the template signal, $Z$  encodes  the scanning strategy   informations and $d$   the time-ordered data acquired by the detector encoding astrophysical signal, noise and (eventually) systematics errors. The linear problem in eq.\ref{eq:pcg} is solved iteratively by means of the Preconditioned Conjugate Gradient (PCG) algorithm.  
 
 Note that this mitigation relies  on how much  the template signal is a good approximation of  the  simulated one $d$. Furthermore, the more   \emph{redundant }  the  scanning strategy, i.e. the more a give line of sight is observed with different scanning orientation,  the higher the signal-to-noise will be in determining the amplitudes $\hat{a}$.  
 
 \subsection{\lb  Calibration systematic simulations} 
 
  We implement  in TOAST the injection of calibration errors and gain drifts as well as the template fitting mitigation\footnote{For further details see \url{https://github.com/hpc4cmb/toast/tree/master/src/toast},
  DOI Zenodo \url{10.5281/zenodo.4270476}}. We then perform simulations for one  \lb  frequency channel,  i.e. 48 polarization sensitive detectors 140 GHz observing for 1 year with the nominal  observing strategy. We adopt the  parameter values for calibration errors and gain drifts  as reported in this note  and inject  both calibration systematics.
  
  The simulated signal encodes: i) the CMB polarized anisotropies (including Solar dipole), ii)  realistic  polarized and unpolarized  Galactic emissions  at the sub-mm wavelengths.  On the other hand,  we built a template signal encoding only unpolarized Galactic emission and the dipole, to  assess  the quality of the mitigation in the presence of an incomplete template signal. 
  
  The output maps are estimated with \texttt{libmadam} map-maker  \citep{Keihanen2010}.  The residuals in  Figure \ref{fig:1} are only due to the systematic errors pre- and post-mitigation, the other set includes  also instrumental  noise. We note that the template fitting procedure is able to reduce the leakage by two orders of magnitude at the  angular power spectrum level indicating that even without a representative template map the \lb scanning strategy is optimized in such a way that we are able to estimate with good signal-to-noise ratio the gain amplitude. Moreover, as expected, this procedure is not affected by instrumental noise. 
  In fact, we do not see any effect in the post-mitigation power spectra with noisy simulations.

 \section{Conclusions}
 
 We  implemented into  TOAST package several   modules  to inject  systematic uncertainties due to beam and calibration. Although not mentioned in this work,  TOAST can also simulate frequency bandpass mismatches, correlated noise  and half-wave plate non-idealities,  making it a  promising  simulation and data analysis  framework for the forthcoming CMB experiments (both ground-based   and space experiments).
 We  propose a mitigation technique that aims at further reducing the leakage induced by gain drifts and calibration errors. The template fitting approach can be thus extended to other kind of systematics effects, e.g. bandpass mismatches, beam imperfections.

\acknowledgments

We acknowledge financial support from the APRA grant: ``Overcoming Systematic Effects in Cosmic Microwave Background Satellite Missions'', Grant number 80NSSC19K0697.  These calculations were performed on the  NERSC supercomputer facility.
\bibliographystyle{plainnat}
\bibliography{references}

\end{document}